# GENERALIZED ESTIMATES FOR THE DENSITY OF OXIDE SCALE IN THE RANGE FROM 0 ºC TO 1300 ºC

*Emmanuil Beygelzimer[1], Yan Beygelzimer[2]*
[1]*OMD-Engineering LLC, Dnipro, Ukraine*
[2]*Donetsk Institute for Physics and Engineering named after O. O. Galkin, National Academy of Sciences, Kyiv, Ukraine*

\*Corresponding author: emmanuilomd@gmail.com, tel.: +380 (50) 368-63-42, 49000, Volodymyr Monomakh Street, 6 off. 303, Dnipro, Ukraine.

**ABSTRACT**

Oxide scale formed on the surface of steel products during high-temperature processes is studied as a composite material, the main solid components of which, in general, are wüstite, magnetite, hematite and metallic iron. To estimate the density of each of these components in the temperature range from 0 °C to 1300 ºC, formulas are proposed that are consistent with the empirical functions of the coefficient of linear thermal expansion, which the author obtained earlier by generalizing data from open sources. The Curie and polymorphic transformation temperatures are included in these generalized formulas as variable parameters, which allows one to take into account the movability of phase transitions due to impurities, crystal lattice defects, particle sizes, cooling rate, and other factors. When specifying the particular values of critical temperatures, the other parameters of the formulas are recalculated automatically. In a particular form, the proposed formulas correspond to the basic values of critical temperatures. According to the calculation examples given, the true (not including pores) density of oxide scale can be about 5200 to 5600 kg·m$^{-3}$, depending on the temperature and percentage of components, whereby a local density minimum may be observed in the region of 570 °C due to eutectoid decomposition of wüstite into magnetite and iron. The proposed methods are recommended for use in mathematical simulation of processing of steel products in the presence of oxide scale on its surface.

**Keywords:** density; wüstite; magnetite; hematite; oxide scale; Curie point

## INTRODUCTION

Thermal and physical properties of surface oxide scale, including its density, are of great importance in the production and processing of steel products. Based on experimental data on the density of rolling mill scale at room temperature [1-4; 5, p. 49-50; 6, p. 102], we can assume that in different technological processes of heating, cooling and deformation of steel the density of oxide scale on its surfaces varies in the range from about 3700 kg·m$^{-3}$ to 5700 kg·m$^{-3}$ depending on its porosity, structural composition and temperature.

At the same time, in mathematical models of such processes, the density of oxide scale and/or individual oxides, that make up its composition, are now taken into account by some constant values that do not depend on the temperature [7-15]. And the values accepted by different authors may differ significantly. For example, in [8, p. 54] the density of wüstite (Fe$_{1-x}$O) is taken equal to 5500 kg·m$^{-3}$, in [12] - 5745 kg·m$^{-3}$, and in [9-11] - 7750 kg·m$^{-3}$.

The assumption of constancy of the density of oxide scale and/or its components is undoubtedly justified in many cases of engineering calculations. However, there are problems, for example, in modeling thermal processes, in which such an approach can lead to large additional errors. To obtain correct computational estimates, it is necessary to take into account that the densities of individual oxides and scale as a whole actually depend on temperature. In this case, it is important that for each component of oxide scale its density estimates are consistent with the data on its other properties. To do this, three conditions should be met: 1) estimates of density and other thermophysical properties should be based on a generalization of different empirical data (rather than one specific study); 2) estimates of density and thermal expansion should be based on the same data (since these properties are physically related to each other); 3) for a given component of oxide scale, the density value should be estimated at the same values of phase transition temperatures as for other thermal physical parameters. The latter is explained by the fact that phase transitions are usually accompanied by a sharp change in properties, and even small difference in the temperatures of a same phase transition taken in estimation of density and, for example, thermal expansion, can lead to inconsistency in the calculated estimates of these properties.

Based on the above, the author set **the goal** to obtain generalized formulas for calculating the density of the main structural components of oxide scale, consistent with the data on their thermal expansion. As the target temperature range the range traditional for mass processes of metal production and processing from 0 ºC to 1300 ºC was taken.

## METHODS

Oxide scale on the surface of steel products is studied as a composite material, the main components of which, in general, are: wüstite FeO (more precisely, Fe$_{1-x}$O, where $x \approx 0.05...0.15$), magnetite Fe$_3$O$_4$, hematite Fe$_2$O$_3$, metallic iron, oxides of alloying elements and gas voids [5, c. 8, 46; 16; 17].

The following temperatures are considered critical: for Fe$_3$O$_4$ and Fe$_2$O$_3$ - the magnetic transition temperatures (Curie points), for iron - the Curie point and the temperature of the polymorphic ($\alpha \leftrightarrow \gamma$) transformation, and for Fe$_{1-x}$O - the temperature of its eutectoid decomposition into magnetite and metallic iron (Chaudron point). The movability ranges and the accepted basic values for these critical temperatures are given in [18].

To estimate the temperature dependence of the density of solid components of oxide scale $\rho(T)$, we used the well-known relationship for anisotropic materials:

$$\rho(T) = \frac{\rho^0}{1 + 3\bar{\alpha}(T - T^0)} \quad (1)$$

where $\rho^0$ is the density of a given component at some initial temperature $T^0$ (i.e. $\rho^0 \equiv \rho(T^0)$); $\bar{\alpha}$ is the *mean* Coefficient of Linear Thermal Expansion (CLTE) of a given component in the interval from $T^0$ to $T$. It was taken into account that formula (1) is valid only for temperature intervals that do not include polymorphic transformations, at which the density of the material changes by leaps.

For consistency purposes formulas for calculating the mean CLTE of the components of oxide scale were obtained by integrating functions for the *true* CLTE, rather than by independently approximating the primary data for the mean coefficient $\bar{\alpha}$. In other words, formulas are based on the integral:

$$\bar{\alpha} = \int_{T^0}^{T} \alpha(T)\, dT \quad (2)$$

where $\alpha(T)$ is a function of the true CLTE of a given component, obtained in [18].

## RESULTS AND DISCUSSION

**Magnetite Fe$_3$O$_4$**

It is known that magnetite exists in only one polymorphic state [19, p. 32]. Therefore, the dependence of magnetite density on temperature in the entire target range from 0 °C to 1300 ° C can be described by a single formula of the form (1):

$$\rho_{Fe3O4} = \frac{\rho^0_{Fe3O4}}{1 + 3\overline{\alpha_{Fe3O4}}(T - 293)} \quad (3)$$

where $\rho_{Fe3O4}$ [kg·m$^{-3}$] and $\rho^0_{Fe3O4}$ [kg·m$^{-3}$] are the density of magnetite at the current temperature $T$ [K] and at 293 K (20 °C), respectively; $\overline{\alpha_{Fe3O4}}$ [K$^{-1}$] is the mean CLTE of magnetite in the temperature range from 293 K to $T$.

Based on known data (**Table 1**), the density of magnetite at 293 K (20 °C) in engineering calculations is recommended to be taken as:

$$\rho^0_{Fe3O4} = 5150 \text{ kg} \cdot \text{m}^{-3} \quad (4)$$

**Table 1.** Data on the density of $Fe_3O_4$ (magnetite) under conditions close to standard one (temperature of 20 °C and pressure of 1 atm)

| Density, kg·m$^{-3}$ | Nature of the data | Reference |
|---|---|---|
| 4175...5215 | Measurements | [20] |
| 4820…5118 | Measurements | [21, p. 59-64] |
| 5000 | Measurements | [22, p. 71] |
| 5010…5060 | Measurements | [23] |
| 5080 | Measurements | [24] |
| 5110…5120 | Calculation | [20] |
| 5170…5180 | Calculation | [25; 26, p. 254] |
| 5201 | Calculation | [21, p. 59; 22, p. 71; 27] |
| 5238 | Calculation | [28] |

The mean CLTE $\overline{\alpha_{Fe3O4}}$ for formula (3) is determined by integrating the functions from [18] for the true CLTE $\alpha_{Fe3O4}(T)$. The final formulas in general form are given in **Table 2** separately for the two intervals separated by the Curie point $T_1$. At the basic value of the Curie point of magnetite $T_1 = 848$ K (575 ºC), these formulas are reduced to a particular form ($\overline{\alpha_{Fe3O4}}$ in [K$^{-1}$], $T$ in [K]):
− in the range of 273 K ≤ $T$ ≤ 848 K

$$\overline{\alpha_{Fe3O4}} = \frac{\Phi_1(T) - 1417.0}{T - 293} \cdot 10^{-6} \tag{5}$$

where $\Phi_1(T)$ is the integral function in the lower temperature range:

$$\Phi_1(T) = -29.571 \cdot T + 19.256 \cdot T^{1.1} + 2000 \cdot e^{-0.005(848-T)} \tag{6}$$

− in the range of 848 K < $T$ ≤ 1573 K

$$\overline{\alpha_{Fe3O4}} = \frac{\Phi_2(T) + 10167}{T - 293} \cdot 10^{-6} \tag{7}$$

where $\Phi_2(T)$ is the integral function in the upper temperature range:

$$\Phi_2(T) = -20.545 \cdot T + 1.3260 \cdot T^{1.4} - 1875 \cdot e^{-0.008(T-848)} \tag{8}$$

The graph of $\overline{\alpha_{Fe3O4}}$, calculated by formulas (5)-(8), in comparison with empirical data is shown in **Fig. 1**.
The results of calculating the magnetite density using formulas (3)-(8), i.e., at the basic value of the Curie point of magnetite $T_1 = 848$ K (575 ºC), are shown as a graph in **Fig. 2**.

**Table 2.** Formulas for calculating the mean CLTE of $Fe_3O_4$ from 293 K (20 °C) to the current temperature $T$ ($T_1$ [K] is the Curie point of magnetite).

| Border of the averaging interval [K] | $273 < T \leq T_1$ | | |
|---|---|---|---|
| Calculation formula [K$^{-1}$] | $\overline{\alpha_{Fe3O4}} = \frac{\Phi_1(T) - \Phi_1(293)}{T - 293} \cdot 10^{-6}$ | | |
| Integral function in the lower temperature range | $\Phi_1(T) = a_0 T + \frac{a_1}{n+1} T^{n+1} + \frac{a_3}{a_4} e^{-a_4(T_1 - T)}$ | | |
| Constants | $n = 0.1$ | $a_3 = 10$ | $a_4 = 0.005$ |
| Coordinates of the reference points | $T_0 = 200$ K | $y_0 = 6.8$ K$^{-1}$ | $y_1 = 22.0$ K$^{-1}$ |
| Coefficients to be calculated | $a_1 = \frac{y_1 - y_0 - a_3(1 - e^{-a_4(T_1 - T_0)})}{T_1^n - T_0^n}$ | | |
| | $a_0 = y_1 - a_1 T_1^n - a_3$ | | |
| Border of the averaging interval [K] | $T_1 < T \leq 1573$ | | |
| Calculation formula [K$^{-1}$] | $\overline{\alpha_{Fe3O4}} = \frac{\Phi_1(T_1) - \Phi_1(293) + \Phi_2(T) - \Phi_2(T_1)}{T - 293} \cdot 10^{-6}$ where $\Phi_1(T)$ is the integral function in the lower temperature range (see above) | | |
| Integral function in the upper temperature range | $\Phi_2(T) = b_0 T + \frac{b_1}{p+1} T^{p+1} - \frac{b_3}{b_4} e^{-b_4(T - T_1)}$ | | |
| Constants | $p = 0.4$ | $b_3 = 15$ | $b_4 = 0.008$ |
| Coordinates of the reference points | $T_2 = 1600$ K | | $y_2 = 15.0$ K$^{-1}$ |
| Coefficients to be calculated | $b_1 = \frac{y_1 - y_2 - b_3(1 - e^{-b_4(T_2 - T_1)})}{T_1^p - T_2^p}$ | | |
| | $b_0 = y_1 - b_1 T_1^p - b_3$ | | |

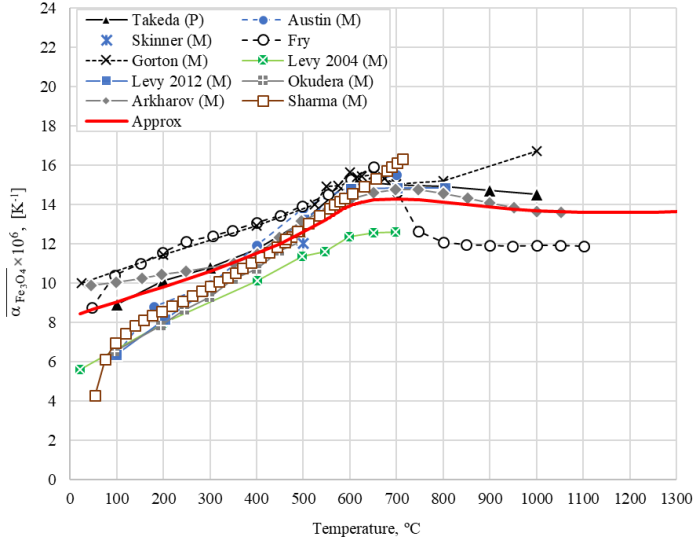

**Fig. 1.** Graph of the mean CLTE of magnetite $\overline{\alpha_{Fe3O4}}$ from 20 °C to a given temperature (values on the horizontal axis) calculated by formulas (3)-(8). Compared to empirical data of Takeda [24], Austin [29], Skinner [30], Fry [31], Gorton [32, p. 279, set 1], Levy [33; 34], Okudera [35], Arkharov [36], Sharma [32, p. 279, sets 2 and 3]. The conventional symbols in the legend of the graph: M - monocrystal, P - polycrystal. The empirical points are conventionally connected by straight lines

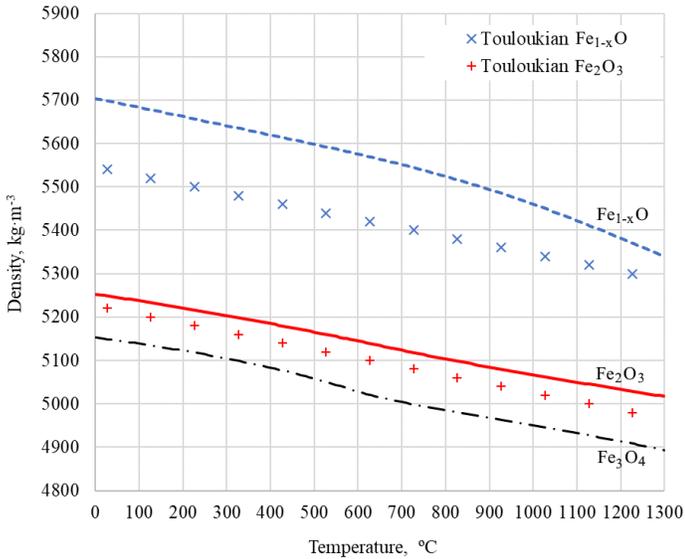

**Fig. 2.** The density of wüstite ($Fe_{1-x}O$), magnetite ($Fe_3O_4$) and hematite ($Fe_2O_3$), calculated by the formulas of this article at basic values of critical temperatures. Compared with the recommendations of Touloukian [37, p. 482, 484].

The sensitivity of magnetite density to the Curie point value is characterized by the data given in **Table 3**. Each row of this table contains $Fe_3O_4$ density values obtained according to the formulas of Table 2 for the same design temperature, but at different values of the Curie point within the real range of its variability from 823 to 900 K (i.e., from 550 to 627 °C - see [18]). It can be seen that the density of $Fe_3O_4$ at the same temperature may be varied by up to 12 kg·m$^{-3}$ depending on the position of the Curie point.

**Table 3.** Calculated effect of the Curie point of magnetite $T_1$ on its density at different design temperatures

| Design temperature [K] ([°C]) | $\rho_{Fe3O4}$ [kg·m$^{-3}$] at different $T_1$ [K] ([°C]) | | | Range [kg·m$^{-3}$] |
|---|---|---|---|---|
| | 823 (550) | 848 (575) | 900 (627) | |
| 773 (500) | 5055 | 5058 | 5063 | 8 |
| 823 (550) | 5039 | 5043 | 5050 | 11 |
| 848 (575) | 5031 | 5035 | 5043 | 12 |
| 900 (627) | 5018 | 5020 | 5026 | 8 |
| 923 (650) | 5012 | 5014 | 5019 | 7 |

**Wüstite $Fe_{1-x}O$**

Since wüstite in the solid phase exists in only one modification [19], its density dependence on temperature can be described by a single formula of the form (1) in the entire target range from 0 °C to 1300 °C:

$$\rho_{FeO} = \frac{\rho_{FeO}^0}{1 + 3\overline{\alpha_{FeO}}(T - 293)} \qquad (9)$$

where $\rho_{FeO}$ [kg·m$^{-3}$] and $\rho_{FeO}^0$ [kg·m$^{-3}$] are the density of wüstite at the current temperature $T$ [K] and at 293 K (20 °C), respectively; $\overline{\alpha_{FeO}}$ [K$^{-1}$] is the mean CLTE of wüstite in the temperature range from 293 K to $T$.

Based on the known data on wüstite density (**Table 4**) the value $\rho_{FeO}^0$ in formula (9) for engineering calculations is recommended to take as:

$$\rho_{FeO}^0 = 5700 \text{ kg} \cdot \text{m}^{-3} \qquad (10)$$

**Table 4.** Data on the density of $Fe_{1-x}O$ (wüstite) under conditions close to standard one (temperature of 20 °C and pressure of 1 atm)

| Density, kg·m$^{-3}$ | Nature of the data | Reference |
|---|---|---|
| 5460 | Measurements | [38] |
| 5590…5950 | Calculation | [39, p. 449] |
| 5613...5728 | Measurements at $x = 0.09…0.055$ | [40] |
| 5652…5742 | Calculation for $x = 0.09…0.055$ | [41, p. 251] |
| 5680 | Measurements at $x = 0.08$ | [42] |
| 5700…6050 | Measurements | [39, p. 449] |
| 5756 | Calculation | [22, p.70] |
| 5841 | Measurements at $x = 0.02$ | [43] |
| 5850…6050 | Measurements | [55] |
| 5865 | Calculation for $x = 0$ | [43] |
| 5870 | Calculation | [44, p. 29] |
| 5900-5990 | Unknown | [19, p. 5] |
| 6000 | Unknown | [25, p. 4-68] |

The formulas for calculating the mean CLTE $\overline{\alpha_{FeO}}$ are obtained by integrating the functions for the true CLTE $\alpha_{FeO}(T)$ [18]. Due to the weak influence of the Chaudron point on the wüstite thermal expansion [18], only a particular form of these formulas for the basic value of the Chaudron point $T_1 = 843$ K (570 °C) is of practical importance ($\overline{\alpha_{FeO}}$ [K$^{-1}$], $T$ [K] – design temperature):

− in the range of 273 K $\leq T \leq$ 843 K

$$\overline{\alpha_{FeO}} = \frac{\Phi_1(T) - 2554.0}{T - 293} \cdot 10^{-6} \qquad (11)$$

where $\Phi_1(T)$ is the integral function in the lower temperature range:

$$\Phi_1(T) = 4.0 \cdot T + 2.3121 \cdot 10^{-2} T^2 - 2.7630 \cdot 10^{-5} T^3 + 1.2487 \cdot 10^{-8} T^4 \qquad (12)$$

− in the range of 843 K $< T \leq$ 1573 K

$$\overline{\alpha_{FeO}} = \frac{\Phi_2(T) - 17787}{T - 293} \cdot 10^{-6} \qquad (13)$$

where $\Phi_2(T)$ is the integral function in the upper temperature range:

$$\Phi_2(T) = 70 \cdot T - 8.5934 \cdot 10^{-2}T^2 + 5.4192 \cdot 10^{-5}T^3 - 1.1121 \cdot 10^{-8}T^4 \quad (14)$$

The graph of $\overline{\alpha_{FeO}}$, calculated by formulas (11)-(14), in comparison with the known empirical data is shown in **Fig. 3**.

The graph of wüstite density calculated by formulas (9)-(14), i.e. at the basic value of Chaudron point $T_1 = 843$ K (570 ºC), is shown in **Fig. 2**. The obtained values are slightly higher than the recommendations of Touloukian [37, p. 482] due to the fact that we accepted a higher density of wüstite at room temperature (5700 kg·m$^{-3}$ vs. 5550 kg·m$^{-3}$).

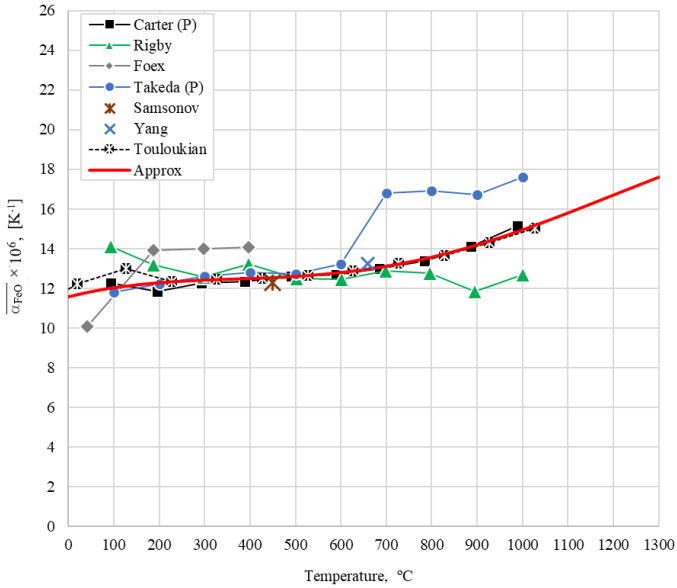

**Fig. 3.** Graph of the mean CLTE of wüstite $\overline{\alpha_{FeO}}$ from 20 ºC to a given temperature (values on the horizontal axis) calculated by formulas (11)-(14). Compared with the empirical data of Carter [45], Rigby [45], Foex [46], Takeda [24], Samsonov [44, p. 131], Yang [47] and the generalized recommendations of Touloukian [37, p. 271]. The empirical points are conventionally connected by straight lines

**Hematite Fe$_2$O$_3$**

Iron oxide Fe$_2$O$_3$ in steel scale in the temperature range from 0 ºC to 1300 º C under normal pressure can be present only in one crystalline form [19]. Therefore, the change in the density of hematite can be described by a formula of the form (1):

$$\rho_{Fe2O3} = \frac{\rho^0_{Fe2O3}}{1 + 3\overline{\alpha_{Fe2O3}}(T - 293)} \quad (15)$$

where $\rho_{Fe2O3}$ [kg·m$^{-3}$] and $\rho^0_{Fe2O3}$ [kg·m$^{-3}$] are the density of hematite at the current temperature $T$ [K] and at 293 K (20 ºC), respectively; $\overline{\alpha_{Fe2O3}}$ [K$^{-1}$] is the mean CLTE of wustite in the temperature range from 293 K to $T$.

Summarizing the known data (**Table 5**), the following value is recommended for engineering calculations as Fe$_2$O$_3$ density at 293 K (20 ºC):

$$\rho^0_{Fe2O3} = 5250 \text{ kg} \cdot \text{m}^{-3} \quad (16)$$

The mean CLTE $\overline{\alpha_{Fe2O3}}$ is determined by integrating functions for the true CLTE $\alpha_{Fe2O3}$ [18]. Formulas for calculating the mean CLTE of hematite $\overline{\alpha_{Fe2O3}}$ [K$^{-1}$] as a function of temperature $T$ [K] are presented in general form in **Table 6**. At the basic value of the Curie point of hematite $T_1 = 950$ K (677 ºC) these dependences take the form:

– in the range of 273 K $\leq T \leq$ 950 K

$$\overline{\alpha_{Fe2O3}} = \frac{\Phi_1(T) - 2082.0}{T - 293} \cdot 10^{-6} \quad (17)$$

where $\Phi_1(T)$ is the integral function in the lower temperature range:

$$\Phi_1(T) = 2.8760 \cdot T + 0.24710 \cdot T^{1,5} + 3.2467 \cdot T^{-1} \quad (18)$$

– in the range of 950 K $< T \leq$ 1573 K

$$\overline{\alpha_{Fe2O3}} = \frac{\Phi_2(T) - 7890.1}{T - 293} \cdot 10^{-6} \quad (19)$$

where $\Phi_2(T)$ is the integral function in the upper temperature range:

$$\Phi_2(T) = 10.259 \cdot T + 988.75 \cdot \ln(T) - 750e^{-0.004(T-950)} \quad (20)$$

**Table 5.** Data on the density of Fe$_2$O$_3$ (hematite) under conditions close to standard one (temperature of 20 °C and pressure of 1 atm)

| Density, kg·m$^{-3}$ | Nature of the data | Reference |
|---|---|---|
| 4690 | Measurements | [24] |
| 4900 | Measurements | [22, p. 71] |
| 5240 | Measurements | [48] |
| 5240 | Unknown | [49] |
| 5250 | Unknown | [25] |
| 5260 | Unknown | [19, p. 5] |
| 5270 | Unknown | [39, p. 449] |
| 5275…5277 | Calculation | [43; 50, p. 67] |
| 5309…5320 | Calculation | [22, p. 71] |

The graph of $\overline{\alpha_{Fe2O3}}$, calculated by formulas (17)-(20), in comparison with the experimental data is shown in **Fig. 4**.

Graph of hematite density, calculated by formulas (16)-(20), i.e. at the basic value of the Curie point $T_1 = 843$ K (570 ºC), is shown in **Fig. 2**. Calculation results are consistent with the recommendations in [37, p. 484].

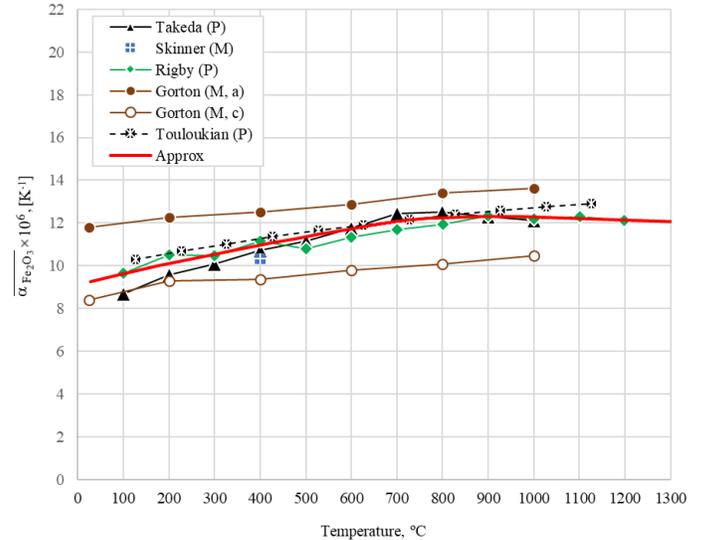

**Fig. 4.** Graph of the mean CLTE of hematite $\overline{\alpha_{Fe2O3}}$ from 20 ºC to a given temperature (values on the horizontal axis) calculated by formulas (17)-(20). Compared with the empirical data of Takeda [24], Skinner [30], Rigby [45], Gorton [32, p. 275] and generalized recommendations of Touloukian [32, p. 274]. The conventional symbols in the legend of the graph: P - polycrystal, M - monocrystal, a and b - indexes of crystal axes. The empirical points are conventionally connected by straight lines.

**Table 6:** Formulas for calculating the mean CLTE of $Fe_2O_3$ from 293 K (20 °C) to the current temperature $T$ ($T_1$ [K] is the Curie point of hematite)

| Border of the averaging interval [K] | $273 \leq T \leq T_1$ | | |
|---|---|---|---|
| Calculation formula [K$^{-1}$] | $\overline{\alpha_{Fe2O3}} = \dfrac{\Phi_1(T) - \Phi_1(293)}{T - 293} \cdot 10^{-6}$ | | |
| Integral function in the lower temperature range | $\Phi_1(T) = a_0 T + \dfrac{a_1}{n+1} T^{n+1} + \dfrac{a_2}{m+1} T^{m+1}$ | | |
| Constants | $n = 0.5$ | | $m = -2.0$ |
| Coordinates of the reference points | $T_0 = 273$ K | $y_0 = 9.0$ K$^{-1}$ | $y_1 = 14.3$ K$^{-1}$ |
| Coefficients to be calculated | $a_0 = \dfrac{y_1 T_0^m + y_0 T_1^n - y_0 T_1^m - y_1 T_0^n}{T_0^n T_1^m - T_1^n T_0^m - T_1^m + T_0^m + T_1^n - T_0^n}$ $a_1 = \dfrac{y_0 T_1^m - y_1 T_0^m - a_0(T_1^m - T_0^m)}{T_0^n T_1^m - T_1^n T_0^m}$ $a_2 = \dfrac{y_0 T_1^n - y_1 T_0^n - a_0(T_1^n - T_0^n)}{T_0^m T_1^n - T_1^m T_0^n}$ | | |
| Border of the averaging interval [K] | $T_1 < T \leq 1573$ | | |
| Calculation formula [K$^{-1}$] | $\overline{\alpha_{Fe2O3}} = \dfrac{\Phi_1(T_1) - \Phi_1(293) + \Phi_2(T) - \Phi_2(T_1)}{T - 293} \cdot 10^{-6}$ where $\Phi_1(T)$ is the integral function in the lower temperature range (see above) | | |
| Integral function in the lower temperature range | $\Phi_2(T) = b_0 T + b_1 \ln(T) - \dfrac{b_3}{b_4} e^{-b_4(T-T_1)}$ | | |
| Constants | $p = -1.0$ | $b_3 = 3.0$ | $b_4 = 0.004$ |
| Coordinates of the reference points | $T_2 = 1600$ K | | $y_2 = 11.1$ K$^{-1}$ |
| Coefficients to be calculated | $b_1 = \dfrac{y_1 - y_2 - b_3(1 - e^{-b_4(T_2-T_1)})}{T_1^p - T_2^p}$ $b_0 = y_1 - b_1 T_1^p - b_3$ | | |

## Iron Fe

In the temperature range of interest from 0 °C to 1300 °C iron undergoes two phase transitions: magnetic at the basic value of the Curie point $T_1 = 1043$ K (770 °C) and polymorphic at the basic temperature under equilibrium conditions $T_2 = 1185$ K (912 °C). The magnetic transition occurs without changing the crystal lattice. During polymorphic transformation, the lattice type and dimensions change from face-centered cubic ($\gamma$-Fe) above $T_2$ to body-centered cubic ($\alpha$-Fe) below $T_2$. Therefore, the dependence of iron density on temperature is described by a piecewise smooth function with a discontinuity at the point $T_2$:

– below the polymorphic transformation temperature: 273 K $\leq T \leq T_2$ ($\alpha$-Fe)

$$\rho_{Fe} = \dfrac{\rho^0_{\alpha-Fe}}{1 + 3\overline{\alpha_{\alpha-Fe}}(T - 293)} \quad (21)$$

– above the polymorphic transformation temperature: $T_2 < T \leq 1573$ K ($\gamma$-Fe)

$$\rho_{Fe} = \dfrac{\rho^{00}_{\gamma-Fe}}{1 + 3\overline{\alpha_{\gamma-Fe}}(T - T_2)} \quad (22)$$

where $T$ [K] is the design temperature; $T_2$ [K] is the temperature of polymorphic transformation of iron; $\rho_{Fe}$ [kg·m$^{-3}$] is the density of iron at the design temperature; $\rho^0_{\alpha-Fe}$ [kg·m$^{-3}$] is the density of $\alpha$-iron at 293 K (20 °C); $\rho^{00}_{\gamma-Fe}$ [kg·m$^{-3}$] is the density of $\gamma$-iron at temperature $T_2$; $\overline{\alpha_{\alpha-Fe}}$ [K$^{-1}$] – mean CLTE of $\alpha$-iron from 293 K to the design temperature $T$; $\overline{\alpha_{\gamma-Fe}}$ [K$^{-1}$] - mean CLTE of $\gamma$-iron from temperature $T_2$ to the design temperature $T$.

The value of iron density at 293 K (20 °C) in formula (21) is recommended to be taken as equal [51, c. 53; 37, c. 443]:

$$\rho^0_{\alpha-Fe} = 7870 \text{ kg} \cdot m^{-3} \quad (23)$$

The mean CLTE of $\alpha$-iron involved in formula (21) was determined on the basis of integrating functions for the true CLTE coefficient of iron $\alpha_{Fe}$ [18] in two temperature intervals: below Curie point ($T_1$) and between Curie and polymorphic transformation points $T_2$. The final formulas in general form (i.e., with varying values of $T_1$ and $T_2$; applicable for $T_1 < T_2$) are given in **Table 7**. With the basic values of the Curie point $T_1 = 1043$ K (770 °C) and the polymorphic transformation temperature $T_2 = 1185$ K (912 °C), these formulas for calculating the mean CLTE of $\alpha$-iron $\overline{\alpha_{\alpha-Fe}}$ [K$^{-1}$] from 293 K (20 °C) to temperature $T$ [K] take the form:

– up to Curie point (273 K $\leq T \leq 1043$ K):

$$\overline{\alpha_{\alpha-Fe}} = \dfrac{\Phi_1(T) - 2252.1}{T - 293} \cdot 10^{-6} \quad (24)$$

where $\Phi_1(T)$ is the integral function in the lower temperature range:

$$\Phi_1(T) = -21 \cdot T + 12.951 \cdot T^{1.14} - 543.40 \cdot e^{-0.013(1043-T)} \quad (25)$$

– between Curie and polymorphic transformation points (1043 K $< T \leq 1185$ K)

$$\overline{\alpha_{\alpha-Fe}} = \dfrac{\Phi_2(T) - 5750.5}{T - 293} \cdot 10^{-6} \quad (26)$$

where $\Phi_2(T)$ is the integral function in the mid-temperature range:

$$\Phi_2(T) = 16.004 \cdot T + 100.08 \cdot e^{-0.05(T-1043)} \qquad (27)$$

The graph of the mean CLTE of α-iron, calculated by formulas (24)-(27), is shown in **Fig. 5**.

**Table 7.** Formulas for the mean CLTE of α-iron from 293 K (20 °C) to $T$ ($T_1$ [K] is the Curie point, $T_2$ [K] is the point of polymorphic transformation)

| Border of the averaging interval [K] | $273\ \text{K} < T \leq T_1$ | | |
|---|---|---|---|
| Calculation formula [K$^{-1}$] | $\overline{\alpha_{\alpha-\text{Fe}}} = \dfrac{\Phi_1(T) - \Phi_1(293)}{T - 293} \cdot 10^{-6}$ | | |
| Integral function in the lower temperature range | $\Phi_1(T) = a_0 T + \dfrac{a_1}{n+1} T^{n+1} + \dfrac{a_3}{a_4} e^{-a_4(T_1-T)}$ | | |
| Constants | $n = 0.14$ | $a_0 = -21.0$ | $a_4 = 0.013$ |
| Coordinates of the reference points | $T_0 = 200$ K | $y_0 = 10.0$ K$^{-1}$ | $y_1 = 11.0$ K$^{-1}$ |
| Coefficients to be calculated | $a_1 = \dfrac{W_1(y_1 - a_0) + a_0 - y_0}{W_1 T_1^n - T_0^n}$ | | |
| | $a_3 = y_1 - a_0 - a_1 T_1^n$ | | |
| Auxiliary parameter | $W_1 = e^{-a_4(T_1 - T_0)}$ | | |
| Border of the averaging interval [K] | $T_1 < T \leq T_2$ | | |
| Calculation formula [K$^{-1}$] | $\overline{\alpha_{\alpha-\text{Fe}}} = \dfrac{\Phi_1(T_1) - \Phi_1(293) + \Phi_2(T) - \Phi_2(T_1)}{T - 293} \cdot 10^{-6}$ where $\Phi_1(T)$ is the Integral function in the lower temperature range (see above) | | |
| Integral function in the lower temperature range | $\Phi_2(T) = b_0 T - \dfrac{b_3}{b_4} e^{-b_4(T-T_1)}$ | | |
| Constants | $b_4 = 0.05$ | | |
| Coordinates of the reference points | $y_2 = 16.0$ K$^{-1}$ | | |
| Coefficients to be calculated | $b_3 = \dfrac{y_1 - y_2}{1 - W_2}$ | | |
| | $b_0 = y_1 - b_3$ | | |
| Auxiliary parameter | $W_2 = e^{-b_4(T_2 - T_1)}$ | | |

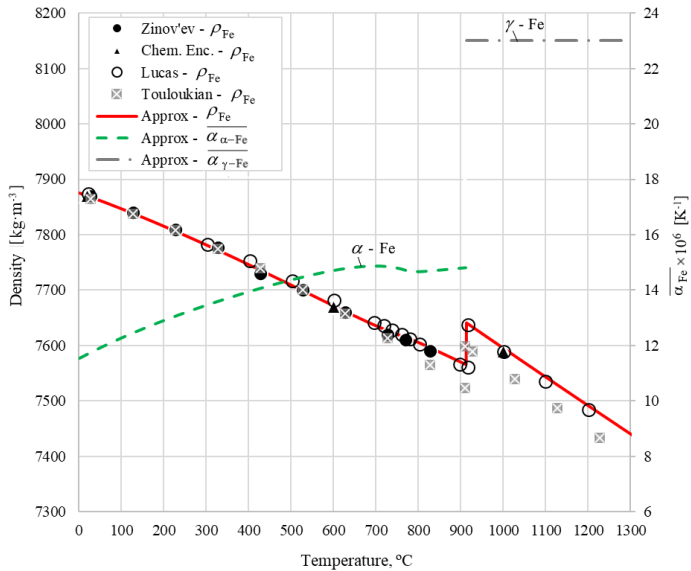

**Figure 5.** The mean CLTE $\overline{\alpha_{\text{Fe}}}$ and density $\rho_{\text{Fe}}$ of iron calculated by the formulas of this article for the basic values of Curie point $T_1 = 1043$ K (770 °C) and polymorphic transformation temperature $T_2 = 1185$ K (912 °C). The density values are compared with the data of Zinov'ev [54, p. 312], Chemical Encyclopedia [26, p. 140], Lucas [51, p. 56] and Touloukian [37, p. 443].

The value of γ-iron density [kg·m$^{-3}$] at the temperature of polymorphic transformation $T_2$ in formula (22) can be calculated as follows,

$$\rho_{\gamma-\text{Fe}}^{00} = k_{\alpha-\gamma} \cdot \rho_{\alpha-\text{Fe}}^{00} \qquad (28)$$

where $k_{\alpha-\gamma}$ is the density "jump" coefficient of α-γ transformation at temperature $T_2$; $\rho_{\alpha-\text{Fe}}^{00}$ [kg·m$^{-3}$] is the density of α-iron at temperature $T_2$.

The "jump" coefficient $k_{\alpha-\gamma}$ can be estimated by the ratio of the volumes of the corresponding atoms:

$$k_{\alpha-\gamma} = \frac{V_\alpha}{V_\gamma} \qquad (29)$$

where $V_\alpha$ and $V_\gamma$ are atomic volume [nm$^3$] of α-Fe (2 atoms per cell) and γ-Fe (4 atoms per cell) at temperature $T_2$,

$$V_\alpha = \frac{1}{2} a_\alpha^3 \qquad (30)$$

$$V_\gamma = \frac{1}{4} a_\gamma^3 \qquad (31)$$

$a_\alpha$ and $a_\gamma$ are lattice parameter [nm] of α-Fe and γ-Fe accordingly at temperature $T_2$. At a base value of $T_2 = 1185$ K (912 °C), the lattice parameters are equal:

$$a_\alpha = 2.8863 \cdot (1 + 15.3 \cdot 10^{-6} \cdot (1185 - 800)) = 2.9033\ \text{nm} \qquad (32)$$

$$a_\gamma = 3.6306 \cdot (1 + 23.0 \cdot 10^{-6} \cdot (1185 - 1000)) = 3.6460\ \text{nm} \qquad (33)$$

where 2.8863 nm is the measured value of the lattice parameter $\alpha$-Fe at 800 K (527 °C); 3.6306 nm is the empirical value of the lattice parameter $\gamma$-Fe at 1000 K (727 °C) (both values from [52], with the value for $\gamma$-Fe obtained by approximating the empirical dependence of the austenite lattice parameter on carbon content); $15.3 \cdot 10^{-6}$ K$^{-1}$ is the calculated value of mean CLTE of $\alpha$-iron in the temperature range from 800 K to 1185 K, $23.0 \cdot 10^{-6}$ K$^{-1}$ is the calculated value of mean CLTE of $\gamma$-iron in the temperature range from 1000 K to 1185 K (both calculated values of mean CLTE were obtained by integrating the corresponding functions of true CLTE of iron [18]).

Substituting (30)-(33) into expression (29), one obtains:

$$k_{\alpha-\gamma} = 1{,}01 \qquad (34)$$

i.e. at the point of polymorphic transformation, the density of $\gamma$-iron is 1% higher than that of $\alpha$-iron (this agrees with the data [53, p. 199]).

The density of $\alpha$-iron at $T_2$ in formula (28) can be calculated as:

$$\rho_{\alpha-\text{Fe}}^{00} = \frac{\rho_{\alpha-\text{Fe}}^0}{1 + 3\overline{\alpha_{\alpha-\text{Fe}}^{00}}(T_2 - 293)} \qquad (35)$$

where $\overline{\alpha_{\alpha-\text{Fe}}^{00}}$ is the mean CLTE of $\alpha$-iron [K$^{-1}$] from 293 K to temperature $T_2$. At basic values of critical temperatures, this coefficient is equal (according to formulas (26)-(27) at $T = 1185$ K):

$$\overline{\alpha_{\alpha-\text{Fe}}^{00}} = 14.81 \cdot 10^{-6} \text{ K}^{-1} \qquad (36)$$

Since the true CLTE of $\gamma$-iron is assumed constant and equal to $23{,}0 \cdot 10^{-6}$ K$^{-1}$ (see [18]), its mean CLTE from the temperature of polymorphic transformation $T_2$ to the current temperature $T$, which is included in formula (22), is also constant:

$$\overline{\alpha_{\gamma-\text{Fe}}} = 23.0 \cdot 10^{-6} \text{ K}^{-1} \qquad (37)$$

Taking into account (28) and (34)-(37), formula (22) for the density of $\gamma$-Fe at the basic values of critical temperatures $T_1$ and $T_2$ can be transformed to the form:
– above the polymorphic transformation temperature: $T_2 < T \leq 1573$ K ($\gamma$-Fe)

$$\rho_{\text{Fe}} = \frac{0.97150 \cdot \rho_{\alpha-\text{Fe}}^0}{1 + 6.9 \cdot 10^{-5}(T - 1185)} \qquad (38)$$

where $\rho_{\alpha-\text{Fe}}^0$ is the density of $\alpha$-iron at 293 K (20 °C) [kg·m$^{-3}$] (see (23)).

The graph of iron density calculated from the above formulas at the basic values of critical temperatures $T_1 = 1043$ K (770 °C) and $T_2 = 1185$ K (912 °C), compared with some data from other sources, is shown in **Fig. 5**. In the range of 800-1300 °C the obtained values of iron density are higher than according to Touloukian's recommendations [37, p. 443] due to the fact that our approximation takes into account the reduction of thermal expansion coefficient in the Curie point area.

Calculations of iron density at different values of critical temperatures using formulas from Table 7 show that the shift of the Curie point within the real range of 1032-1046 K (759-773 °C) practically does not influence the density of iron, while the mobility of polymorphic transformation may result in a change of iron density at a fixed temperature by up to 75 kg·m$^{-3}$ due to a jump change in the crystal lattice parameters.

**Oxide scale as a whole**

The true (i.e., excluding pores) density of oxide scale as a multicomponent material can be represented as:

$$\rho_{sc} = \frac{m_{sc}}{V_{sc}} = \frac{\rho_1 V_1 + \rho_2 V_2 + \cdots}{V_{sc}} = \rho_1 \psi_1 + \rho_2 \psi_2 + \cdots \qquad (39)$$

where $\rho_{sc}$ is the true density of oxide scale; $m_{sc}$ is the mass of a given amount of scale, which without pores would occupy the volume $V_{sc}$; $\rho_1, \rho_2, \ldots$ - the density of the 1-st, 2-nd and other components accordingly; $V_1, V_2, \ldots$ - volume of each component in a given amount of scale (also without considering the pores); $\psi_1, \psi_2, \ldots$ - volume fractions of each component:

$$\psi_1 = \frac{V_1}{V_{sc}}, \quad \psi_2 = \frac{V_2}{V_{sc}}, \ldots \qquad (40)$$

On this basis, the true density of oxide scale can be calculated by the formula:

$$\rho_{sc} = \psi_{\text{FeO}} \cdot \rho_{\text{FeO}} + \psi_{\text{Fe3O4}} \cdot \rho_{\text{Fe3O4}} + \psi_{\text{Fe2O3}} \cdot \rho_{\text{Fe2O3}} + \psi_{\text{Fe}} \cdot \rho_{\text{Fe}} + \psi_{x\text{O}} \cdot \rho_{x\text{O}} \qquad (41)$$

where the symbol $\rho$ with an index denotes the density of the corresponding component ($x$O means the oxide of the alloying element), and $\psi$ with an index denotes its fraction in the "pure" (i.e. without considering the pores) volume of the scale, thus:

$$\psi_{\text{FeO}} + \psi_{\text{Fe3O4}} + \psi_{\text{Fe2O3}} + \psi_{\text{Fe}} + \psi_{x\text{O}} = 1 \qquad (42)$$

As an example, **Fig. 6** shows graphs of $\rho_{sc}$, calculated by the formula (41) for four hypothetical compositions of oxide scale according to **Table 8-9**. From the graphs we can see that the range of possible changes in the true density of oxide scale is approximately from 5200 to 5600 kg·m$^{-3}$ depending on the temperature and the percentage composition of components. It is noteworthy that in the case of eutectoid decomposition of wüstite (composition No. 4) the local minimum of scale density may be observed in the vicinity of Chaudron point (basic value of 570 °C), but not at the maximum temperature.

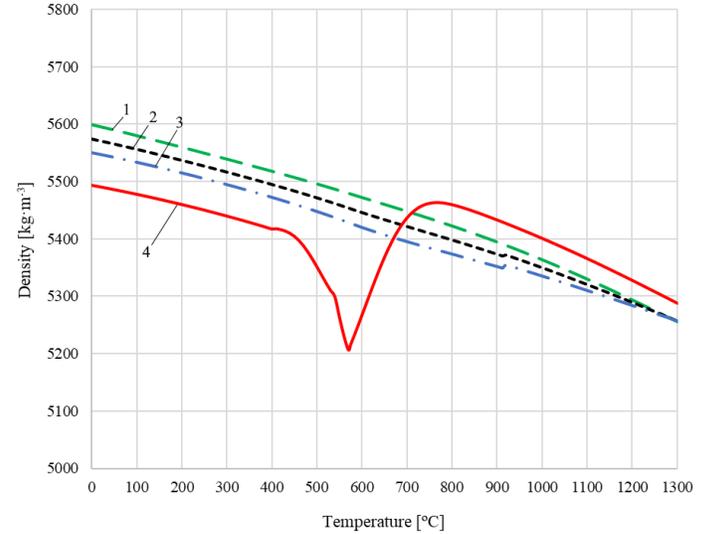

**Fig. 6.** Graphs of true scale density $\rho_{sc}$, calculated by the formula (41) through the density of its structural components at the basic values of critical temperatures. Numbers at the curves are the numbers of scale compositions according to Tables 8-9.

**Table 8.** Constant compositions of scale No. 1-3, adopted for the graphs in Fig. 6.

| Composition | Component volume fraction $\psi$ | | | |
|---|---|---|---|---|
| | Fe$_{1-x}$O | Fe$_3$O$_4$ | Fe$_2$O$_3$ | Fe |
| 1 | 0,8 | 0,15 | 0,05 | 0 |
| 2 | 0,5 | 0,35 | 0,1 | 0,05 |
| 3 | 0,2 | 0,55 | 0,15 | 0,1 |

**Table 9.** Variable composition of scale No. 4, adopted for the graph in Fig. 6 (more details - in [18])

| Temperature, °C | Component volume fraction $\psi$ | | | |
|---|---|---|---|---|
| | Fe$_{1-x}$O | Fe$_3$O$_4$ | Fe$_2$O$_3$ | Fe |
| 1300 | 0,880 | 0,100 | 0,020 | 0 |
| 1000 | 0,880 | 0,100 | 0,020 | 0 |
| 900 | 0,880 | 0,100 | 0,020 | 0 |
| 800 | 0,875 | 0,100 | 0,025 | 0 |
| 700 | 0,779 | 0,166 | 0,055 | 0 |
| 600 | 0,416 | 0,489 | 0,095 | 0 |
| 570 | 0,286 | 0,596 | 0,118 | 0 |
| 500 | 0,068 | 0,749 | 0,090 | 0,093 |
| 400 | 0 | 0,788 | 0,090 | 0,122 |
| 300 | 0 | 0,788 | 0,090 | 0,122 |
| 100 | 0 | 0,788 | 0,090 | 0,122 |

The dependence of oxide scale density on its porosity is accounted for this way [5, c. 48]:

$$\rho'_{sc} = \frac{\rho_{sc}}{1 - \eta} \qquad (43)$$

where $\rho'_{sc}$ is the apparent (including pores) density of scale; $\eta$ is the porosity of oxide scale, which is the ratio of the pore volume to the total volume of scale, including pores, $\rho_{sc}$ is the true (excluding pores) density of scale by the formula (41).

**CONCLUSION**

Temperature dependences of the mean coefficient of linear thermal expansion (CLTE) of the main components of oxide scale (wüstite $Fe_{1-x}O$, magnetite $Fe_3O_4$, hematite $Fe_2O_3$ and metallic iron) were obtained by integrating the previously proposed [1] functions of the true CLTE. The entire range from 0 °C to 1300 °C is divided into intervals by critical temperatures, which are considered the points of magnetic transitions, polymorphic transformation (for Fe) and thermodynamic stability boundary (for $Fe_{1-x}O$). In general form, these critical temperatures enter the formulas as varying parameters. In particular form the formulas correspond to the basic values of critical temperatures.

The obtained formulas for the mean CLTE are used to estimate the density dependence on temperature for each component of the oxide scale. For magnetite, the Curie temperature variation in the 550-627ºC range can lead to density changes of the order of 10 kg·m$^{-3}$ in the vicinity of this interval, whereas for wüstite and hematite the possible shift in magnetic transition temperature has almost no effect on their density. For iron, the Curie point shift also has virtually no effect on its density at a particular temperature, but the mobility of the polymorphic ($\alpha \leftrightarrow \gamma$) transformation may be led to a change in density at a fixed temperature by up to 75 kg·m$^{-3}$ due to a jump change in the crystal lattice parameters.

To calculate the true density of oxide scale as a whole, the sum of the products of the density of its components on their volume fractions used. Calculation examples show that when cooling the steel product in the technological process, the range of possible changes in the true (excluding pores) density of scale is approximately from 5200 to 5600 kg·m$^{-3}$, with a local minimum may be observed in the region of 570 °C due to eutectoid decay of wüstite to magnetite and free iron. The resulting formulas are recommended for use in mathematical simulation of the production and processing of steel products in the presence of surface scale.